\documentclass[a4paper, 11pt]{article}
\usepackage[margin=2.5cm]{geometry}
\usepackage[cmex10]{amsmath}
\usepackage{amssymb}
\usepackage{amsthm}
\usepackage{array}
\usepackage{accents}
\makeatletter
  \def\widebar{\accentset{{\cc@style\underline{\mskip10mu}}}}
  \def\wideubar{\underaccent{{\cc@style\underline{\mskip10mu}}}}
\makeatother
\usepackage{graphicx}
\graphicspath{{./figs/}}
\usepackage{cite}
\usepackage[labelformat=simple]{subcaption}

\usepackage[dvipdfmx]{hyperref}

\newcommand{\R}{\mathbb{R}}
\newcommand{\calY}{\mathcal{Y}}
\newcommand{\bY}{\widebar \calY}
\newcommand{\uY}{\wideubar \calY}
\newcommand{\calA}{\mathcal{A}}
\newcommand{\calC}{\mathcal{C}}
\newcommand{\lambdaAS}{\lambda_{A^*}\!}
\newcommand{\lambdaA}{\lambda_{A}}
\newcommand{\Rnec}{R_{\textnormal{nec}}}

\newcommand{\Nnece}{N_{\textnormal{nec}}^{(\textnormal{e})}}
\newcommand{\Nneco}{N_{\textnormal{nec}}^{(\textnormal{o})}}

\newcommand{\hN}{\hat N}

\theoremstyle{definition}
\newtheorem{thm}{Theorem}
\newtheorem{assum}{Assumption}
\newtheorem{lem}{Lemma}
\newtheorem{rem}{Remark}
\newtheorem{cor}{Corollary}
\newtheorem{exmp}{Example}
\makeatletter
 \newenvironment{pf}[1][\proofname]{\par\pushQED{\qed}
 \normalfont\topsep6\p@\@plus6\p@\relax\trivlist\item[\hskip\labelsep\bfseries#1\@addpunct{.}]
 \ignorespaces}{\popQED\endtrivlist\@endpefalse}
\makeatother

\newcommand{\fref}[1]{Fig.~\ref{#1}}

\newcommand{\IEEEJAC}{{IEEE} Trans. Autom. Control}
\newcommand{\IEEEJPROC}{Proc. {IEEE}}

\newcommand{\IJRN}{Int. J. Robust Nonlin. Control}
\newcommand{\LAA}{Linear Algebra Appl.}

\newcommand{\SIAMCO}{SIAM J. Control Optim.}
\newcommand{\SysCL}{Syst. Control Lett.}

\title{Minimum data rate for stabilization of linear systems\\
 with parametric uncertainties%
\thanks{%
This work was supported in part by the Ministry of Education, Culture, Sports,
Science and Technology, Japan, under Grant-in-Aid for Scientific Research Grant
No.~23760385 and by the Aihara Project, the FIRST program from JSPS,
initiated by CSTP.
}}

\author{Kunihisa Okano%
\thanks{%
K.~Okano is with the Department of Electrical and Computer Engineering,
University of California,
Santa Barbara, CA 93106, U.S.A. and is a JSPS Research Fellow.
Email: {\tt \small kokano@ece.ucsb.edu}.}
~and Hideaki Ishii%
\thanks{%
H.~Ishii is with the Department of Computational Intelligence and Systems Science,
Tokyo Institute of Technology,
Yokohama, 226-8502, Japan.
Email: {\tt \small ishii@dis.titech.ac.jp}.
}}

\begin{document}
\maketitle

\begin{abstract}
We study a stabilization problem of linear uncertain systems with parametric
uncertainties via feedback control over data-rate-constrained channels.
The objective is to find the limitation on the amount of information that
must be conveyed through the channels for achieving stabilization and
in particular how the plant uncertainties affect it.
We derive a necessary condition and a sufficient condition for stabilizing
the closed-loop system.
These conditions provide limitations in the form of bounds on
data rate and magnitude of uncertainty on plant parameters.
The bounds are characterized by the product of the poles of the nominal
plant and are less conservative than those known in the literature.
In the course of deriving these results, a new class of nonuniform quantizers
is found to be effective in reducing the required data rate.
For scalar plants, these quantizers are shown to minimize the required data rate,
and the obtained conditions become tight.
\end{abstract}

\section{Introduction}\label{sec,intro}
In this paper, we study stabilization of uncertain discrete-time linear systems
over communication channels.
Due to the use of channels, the amount of information transmitted through
a channel at a time step, or the data rate, is limited to a finite number.
In real systems, the rates are finite because of the limitations on bandwidths
and resolutions of sensors.
As a consequence the controller may not have access to the exact plant
states but only their approximated values are given at each step.
Clearly, such communication constraints may be harmful and can cause degradation
in control performance.
In the pioneering work of \cite{Wong1999}, it has been shown that there
exists a critical value in the data rate to stabilize a feedback system
over channels, and moreover the value depends only on the product of the
unstable poles of the open-loop system.
This result has motivated researchers to derive general minimum data rate
theorems \cite{Tatikonda2004} and to extend them to various problems including
stabilization of stochastic systems \cite{Nair2004} and nonlinear systems \cite{Liberzon2005},
control with shared channels among multiple nodes \cite{Nesic2009},
and time-varying data rate constraints \cite{Minero2009}
(see also the survey paper \cite{Nair2007}).
Moreover, it is interesting that several works have pointed out that notions and tools
in information theory are useful in the analysis of data rate limited control problems.
In \cite{Tatikonda2004b}, rate distortion theory is employed to deal with the
linear quadratic Gaussian problem over a channel.
In \cite{Shingin2012}, an entropy-based approach has been established to
study performance limitations on disturbance rejection.

The main feature of this paper is that we take account of uncertainties in
plant models.
Control of uncertain networked systems has been discussed in several recent
works.
In \cite{Phat2004}, linear time-invariant systems with norm bounded
uncertainties are considered, while in \cite{Martins2006} scalar systems
(state variables are scalars) with nonlinear terms, stochastic
uncertainties, and disturbances are dealt with.
Though sufficient conditions on data rates are obtained, these results
do not characterize the minimum rate.
Another related problem in networked control is stabilization based on
the so-called logarithmic quantizers \cite{Elia2001, Fu2005, Tsumura2009};
uncertain plants have been studied in \cite{Fu2010} and \cite{Hayakawa2009}
from robust control and adaptive control viewpoints, respectively.

The focus of our study is to derive limitations on the data rates necessary
for stabilization of uncertain systems. %
In particular, the limitation is characterized by the level of plant instability
as in the data rate theorems mentioned above and is expressed in terms
of data rates and the uncertainty bounds on
plant parameters.
Under the presence of uncertainties, in general, this is a difficult problem.
The reason is that the combination of plant uncertainties and the nonlinearity
in the system due to quantization complicate the analysis of state evolutions.
In this paper, to overcome such difficulties, we formulate the problem
based on two ideas as follows.

First, we assume that the plant is a single-input and single-output system,
and its uncertainties are parametric.
In the analysis of such a system, the product of poles of the nominal plant
can be expressed as a single parameter, which corresponds to the constant
term in the denominator polynomial of the transfer function;
for the case of known plants, this viewpoint has been proposed in \cite{You2010}.
The coefficients of the polynomial lie somewhere within known bounds, and in this sense
the plant is parametrically uncertain.
In the context of robust control, parametrically uncertain systems have
been extensively studied (see, e.g., \cite{Barmish1994, Bhattacharyya1995}).
The celebrated Kharitonov's theorem \cite{Kharitonov1979} provides an exact
condition for robust stability for continuous-time systems,
though its extensions to discrete-time systems are somewhat limited.

The second idea to tackle the uncertain systems case is the introduction
of some structure into the controller by imposing restriction on the state
estimation scheme.
It is important to note that this controller class includes those that
have appeared in minimum data rate results for known plants for derivation
of sufficient conditions \cite{Tatikonda2004, Tatikonda2004a, You2010}.
Therefore, when specialized to the case without uncertainties, our results
coincide with those in such previous works. %
An interesting aspect in the uncertain case is that the quantizer used
in the encoder should not be restricted to the conventional uniform quantizers
as in \cite{Phat2004, Martins2006, Okano2012}.
We propose a new quantizer, which is in fact designed to compensate plant
uncertainties and is capable of reducing the required data rate.
Indeed, for scalar plants, this quantizer becomes optimal in the sense that
it minimizes the required data rate.
This quantizer is a piecewise constant function whose step width shrinks
as the input becomes larger in magnitude.
In the special case of known plants, it becomes uniform, which supports
the use of uniform quantizers in \cite{You2010}.

The paper is organized as follows.
In Section \ref{sec,problem}, we describe the setup of the networked
control system and formally state the stabilization problem that we consider.
Next, in Section \ref{sec,nec}, we present our first main result, which
characterizes the minimum data rate for uncertain plants.
Sections \ref{sec,gnec} and \ref{sec,opt_q} are devoted to provide key lemmas
and the proof of the first main result;
in Section \ref{sec,gnec}, a lower bound on the expansion rate of the
state estimations set is analyzed and the optimal quantizer which can
achieve the lowest expansion rate is provided in Section \ref{sec,opt_q}.
In Section \ref{sec,suf}, a sufficient bound on the data rate is obtained
by constructing a specific control scheme.
Tightness of the presented limitations is discussed in Section \ref{sec,scalar}.
We show the cases that the necessary condition and the sufficient condition
become tight and less conservative than those presented in the existing results.
In Sections \ref{sec,average} and \ref{sec,averagesuf}, we generalize the
control and communication scheme from a static one to a time-varying one
and extend the previous results.
Finally, we conclude the paper in Section \ref{sec,conclusion}.

The material of this paper was presented in \cite{Okano2012b} in
a preliminary form.
In this paper, we denote $\log_2(\cdot)$ simply as $\log(\cdot)$.

\section{Problem Formulation}\label{sec,problem}
\begin{figure}[t]
 \centering
 \includegraphics[scale=1.0]{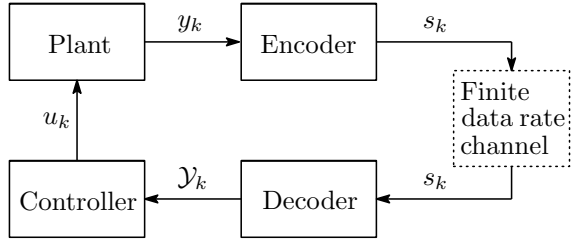}
 \caption{Networked control system.}
 \label{fig,system}
\end{figure}
In this section, we state the setup of the control systems and formulate
the problem to be solved in this paper.

We consider stabilization of a single-input single-output discrete-time system
which has a communication channel at the side of the plant output as depicted
in \fref{fig,system}.
At time $k$, the encoder observes the plant output $y_k\in\R$ and quantizes it.
The quantized signal $s_k\in\Sigma_N$ is transmitted to the decoder through
the finite data rate channel.
Here, the set $\Sigma_N$ represents all possible outputs of the encoder
and contains $N$ symbols.
Thus, the required data rate is expressed as $R:=\log N$ [bits/sample].
From the received signal $s_k$, the decoder computes the interval $\calY_k\subset \R$,
which is an estimate of $y_k$.
Finally, using the past and current estimates, the controller provides
a control input $u_k\in\R$.

In what follows, we describe the details of each component in the system.
The plant is an $n$-dimensional autoregressive system
whose parameters are uncertain\footnote{
The results in this paper can be extended to the case where
the plant is a time-varying ARX model as
$y_{k+1}=\sum^{n_y}_{i=1}a_{i,k}y_{k-i+1}+\sum^{n_u}_{i=1}b_iu_{k-i+1}$,
where $b_1,\dots,b_{n_u}$ are known parameters and $b_1\neq0$.
In \cite{Nair2000}, a related class of plants is studied for limited
data rate control.}:
\begin{align}
  y_{k+1}=a_{1}y_k+a_{2}y_{k-1}+\cdots+a_{n}y_{k-n+1}+u_k.\label{AR}
\end{align}
Here, for the initial values $y_k$, $k=-n+1, -n+2, \dots, 0$, there exist known
bounds $Y_k>0$ as $|y_k|\leq Y_k$, and the input $u_k$ is set to $0$ for $k<0$.
Each uncertain parameter $a_{i}$ is represented by the nominal value $a_i^*$
and the width $\epsilon_i\geq 0$ of the perturbation as
\begin{gather}
 a_{i}\in\calA_i:=\left[a_i^*-\epsilon_i, a_i^*+\epsilon_i\right]\;
 \text{for }i=1,2,\dots,n.\label{uncertainty}
\end{gather}
Let $A^*$ represent the set of parameters $a_i$ where each parameter
is the nominal one, and let $\lambdaAS$ be the product of the poles of the
plant with the parameters $A^*$.

The encoder quantizes the plant output $y_k$ into the $N$-alphabet signal
$s_k\in\Sigma_N$, where $\Sigma_N:=\{1,2,\dots,N\}$.
The input range of the encoder is centered at the origin and the width
is defined by a scaling parameter $\sigma_k>0$. 
In particular, the output $s_k$ of the encoder is given as
\begin{align}
 s_k=q_N\left(\frac{y_k}{\sigma_k}\right),\label{def,s}
\end{align}
where $q_N(\cdot)$ is a static $N$-level quantizer whose input range is $[-1/2,1/2]$.
In the quantizer $q_N$, it is assumed that boundaries of the quantization
cells are symmetric about the origin.
This assumption is introduced to avoid complexity in the analysis regarding
how large the state estimation sets generated by quantized information become.
By its symmetry, the quantizer $q_N$ is characterized by the set of boundary points
$h_{l}\geq0$, $l\in\left\{0,1,\dots,\lceil N/2\rceil\right\}$,
of nonnegative quantization cells.
Here, $\lceil\cdot\rceil$ is the ceiling function.
These points must satisfy
\begin{align}
 h_0=0,\, h_{\lceil N/2\rceil}=\frac{1}{2},\, h_l<h_{l+1}.\label{quantizeredge}
\end{align}
The origin $h_0$ is a boundary only when the number $N$ of quantization cells is
even.
However, for simplicity, we use the same notation above even if $N$ is odd.
Denote the quantization cells determined by $\{h_l\}_l$, from left to right,
by $\calC_i$, $i=1,2,\cdots,N$:

(i) If $N$ is odd, then
\begin{align}
 \calC_i:=\begin{cases}
	  [-h_{\lceil N/2\rceil-i+1},-h_{\lceil N/2\rceil-i}) & \text{if } i\in\{1,2,\cdots,\lceil\frac{N}{2}\rceil-1\},\\
	  [-h_{1},h_{1}) & \text{if } i=\lceil\frac{N}{2}\rceil,\\
	  [h_{i-\lceil N/2\rceil},h_{i+1-\lceil N/2\rceil}) & \text{if } i\in\{\lceil\frac{N}{2}\rceil+1,\lceil\frac{N}{2}\rceil+2,\cdots,N-1\},\\
	  [h_{\lceil N/2\rceil-1},h_{\lceil N/2\rceil}] & \text{if } i=N.
	 \end{cases}\label{def,cell_odd}
\end{align}

(ii) If $N$ is even, then
\begin{align}
 \calC_i:=\begin{cases}
	  [-h_{N/2-i+1},-h_{N/2-i}) & \text{if } i\in\{1,2,\cdots,\frac{N}{2}\},\\
	  [h_{i-1-N/2},h_{i-N/2})   & \text{if } i\in\{\frac{N}{2}+1,\frac{N}{2}+2,\cdots,N-1\},\\
	  [h_{N/2-1},h_{N/2}] & \text{if } i=N.
	 \end{cases}\label{def,cell_even}
\end{align}
Then, for a given set of boundaries $\{h_l\}_l$ of the quantizer $q_N$
and consequently the quantization cells (\ref{def,cell_odd}) or (\ref{def,cell_even}),
we define the outputs of the quantizer as follows:
\begin{align}
 q_N(y):=i \quad\text{if } y\in \calC_i \text{ for } i=1,2,\cdots,N.\label{def,q}
\end{align}

The decoder converts the received signal $s_k$ to the interval
$\calY_k\subset\R$,
which provides an estimate of the set in which the plant
output $y_k$ should be contained.
Formally, $\calY_k$ is defined as the interval
corresponding to the quantization cell that $y_k$ fell in, i.e.,
\begin{align}
 \calY_k:= \sigma_k\calC_{s_k}. \label{def,calY}
\end{align}

The controller provides the control input $u_k$ based on the estimates
$\calY_{k-n+1},\dots,\calY_{k}$ as
\begin{align}
 u_k=\sum^n_{i=1}f_{i,k}\left(\calY_{k-i+1}\right),\label{controller}
\end{align}
where $f_{i,k}(\cdot)$ is an arbitrary map from an interval on $\R$ to a real number.

We remark that the scaling parameter $\sigma_k$ should be large enough
to cover all possible inputs to the encoder. Otherwise, the quantizer may be
saturated, in which case we lose track of the plant output $y_k$.
On the other hand, if we take $\sigma_k$ large, the quantization error also
becomes large. Moreover, to achieve stabilization of the system, $\sigma_k$
should decay to zero gradually.

We determine the scaling parameter $\sigma_k$ as follows.
At time $k$, the encoder and the decoder predict the next plant output
$y_{k+1}$ based on the observed $\calY_0,\dots,\calY_k$.
Let $\calY_{k+1}^-\subset\R$ be the set of all possible outputs
$y_{k+1}$ of the uncertain system (\ref{AR}).
Then the scaling parameter $\sigma_{k+1}$ is chosen such that
\begin{align}
 \sigma_{k+1}\geq\mu(\calY_{k+1}^-),\label{sigma_ineq}
\end{align}
where $\mu(\cdot)$ denotes the Lebesgue measure on $\R$.

The prediction set of the plant output $y_{k+1}$ constructed at time $k$
is defined as follows:
\begin{align}
  \calY^-_{k+1}:=\{&a_1'y_k'+\cdots+a_n'y_{k-n+1}'\!:
  a_1'\in\calA_1,\dots, a_n'\in\calA_n,\notag\\
  &y_k'\in\calY_{k}, \dots, y_{k-n+1}'\in\calY_{k-n+1}
  \}.\label{def,calY^-}
\end{align}
Under this definition, our prediction strategy is to use the information
regarding $y_k,\dots,y_{k-n+1}$ independently such that $y_{k-i+1}\in\calY_{k-i+1}$
for each $i=1,2,\dots,n$, where $\calY_{k-i+1}$ is the interval received
by the decoder at $k-i+1$.
Then, clearly, $\mu(\calY_{k+1}^-)$ is large enough to include $y_{k+1}$,
and is computable on both sides of the channel.

The control objective is to robustly stabilize the networked control
system depicted in \fref{fig,system} for all possible parameters
within the bounds in (\ref{uncertainty}).

The problem setup is particularly affected by the consideration
of uncertain plants.
To overcome the difficulties due to the uncertainty, we have introduced some structures in the
plant as well as the controller.
In the plant (\ref{AR}), the product of the poles is represented as a single
parameter $a_{n}$.
As we mentioned  in Section \ref{sec,intro}, the product of the poles plays
an important role to describe the bounds on the data rate.
Since our objective is to characterize the bounds by the product of
the poles, the simple expression of the key parameter helps to
reduce the complexity in the analysis.

Similarly, the classes of controllers (\ref{controller}) and prediction sets
(\ref{def,calY^-}) are employed to pursue an analytical approach and,
in particular, to obtain necessary limitations in an explicit formula.
Here, we use the information regarding $y_{k-n+1},\dots,y_{k}$ independently.
This may make the state estimation somewhat conservative.
On the other hand, in related works, e.g., \cite{Tatikonda2004, Nair2004, You2010},
the controller can be an arbitrary causal function.
If we use a more general controller or a prediction method that allows us to
look at the correlations among them, then the estimation sets
$\calY_{k-n+1},\dots,\calY_{k-1}$ from times before the current time $k$
may be updated so that they shrink in size.
As a result, the system can be stabilized under a smaller data rate compared
with the case employing (\ref{controller}) and (\ref{def,calY^-}).
We note that it may be possible to minimize the state estimation sets numerically \cite{Rohn1989};
however, in the case of uncertain plants, it is difficult to do this analytically.

Here, regarding the set $\calY_{k+1}^-$, we introduce a useful lemma,
which will be referred to in the following sections.
\begin{lem}\label{lem,BM}
 The prediction set $\calY_{k+1}^-$ defined in (\ref{def,calY^-})
satisfies the following equality:
 \begin{align}
 \mu(\calY^-_{k+1})
 =\sum_{i=1}^n\mu\left(\calA_i\calY_{k-i+1}\right),\label{BM}
\end{align}
where $\calA_i\calY_{k-i+1} := \left\{a'y' : a'\in\calA_i,\,y'\in\calY_{k-i+1}\right\}$
for $i=1,2,\dots,n$.
\end{lem}

\begin{pf}
 \begin{sloppypar}
 By applying the Brunn-Minkowski inequality \cite{Cover2006} to (\ref{def,calY^-}),
 we have $\mu(\calY^-_{k+1})\geq\sum_{i=1}^n\mu\left(\calA_i\calY_{k-i+1}\right)$.
 Furthermore, the equality holds since $\calA_i\calY_{k-i+1}$, $i=1,\dots,n$,
 are connected intervals in $\R$ by the definitions in (\ref{uncertainty})
 and (\ref{def,calY}).
 \end{sloppypar}
\end{pf}

\section{Minimum Data Rate: Fixed Data Rate Case}\label{sec,nec}
In this section, we consider the quantizer (\ref{def,q}) using a fixed number
$N$ of quantization cells.
Under the setup, we present a lower bound on the data rate $R$
as a necessary condition for the system to be stable.
The bound is expressed by the level of instability and uncertainty in the plant.
It is important to note that the previous works \cite{Phat2004, Martins2006, Fu2010}
dealing with uncertainties have studied only sufficient conditions
which contain some conservativeness.

Before providing the result, we introduce an assumption regarding
the poles of the plant.
\begin{assum}\label{asm,instability}
The product of the poles of the plant is greater than $1$ for all
possible parameters in (\ref{uncertainty}), i.e.,
\begin{align}
 |a_n^*|-\epsilon_n>1.\label{a-e>1}
\end{align}
\end{assum}

\begin{rem}
This assumption is required in Sections \ref{sec,nec}, \ref{sec,gnec}, and \ref{sec,average},
where we pursue characterization of necessary data rates for stability,
since our objective is to describe the data rate limitation by the product of poles
$\lambdaAS$.
For cases of known plants \cite{Nair2004, Tatikonda2004, Tatikonda2004a}, %
the data rate bounds are expressed only by the poles outside the unit circle.
In these works, the states corresponding to the stable poles
have been omitted by applying transformation of the state coordinate.
However, if the plant parameters contain uncertainties, we can not make
such a transformation.
If the plant has any stable modes, the limitation on the data rate
given below necessarily becomes loose.
\end{rem}

To describe the main result in this section, we introduce the following notation:
\begin{align}
  r:=\frac{|\lambdaAS|-\epsilon_n}{|\lambdaAS|+\epsilon_n}.\label{def,nu,r}
\end{align}

We are now ready to present the necessity result for the static data rate case.
\begin{thm}\label{th,nec}
 Under Assumption~\ref{asm,instability}, if the system in \fref{fig,system}
is stable with the static quantizer (\ref{def,q}), then it holds that
\begin{align}
 &R>\Rnec:=
 \begin{cases}
  \log\frac{\log(1-\epsilon_n)^2}{\log r} &\text{if }\epsilon_n>0,\\
  \log|\lambdaAS|                         &\text{if }\epsilon_n=0,
 \end{cases}\label{nec,R}\\
 &0\leq \epsilon_n<1\label{nec,e}.
\end{align}
\end{thm}

One can confirm that the lower bound $\Rnec$ on data rate is monotonically
increasing with respect to $|\lambdaAS|$ and $\epsilon_n$.
Thus, more unstable dynamics or more uncertainty in the plant will result
in higher requirement in communication with a larger data rate.
We remark that there is no gap between the two expressions in (\ref{nec,R})
since $\Rnec$ is right continuous with respect to $\epsilon_n$ at $\epsilon_n=0$.

For the special case where no plant uncertainty, i.e., $\epsilon_i=0$ for
$i=1,\dots,n$, the bound $\Rnec$ takes the well-known form $\log|\lambdaAS|$.
In addition, if all poles lie outside of the unit circle, this bound coincides
with those given in \cite{Tatikonda2004, Nair2004}.
It is interesting to note that for an uncertain plant with $\epsilon_n>0$
and $\Rnec>1$, the following inequality holds:
\begin{align}
 \Rnec>\max_{\lambdaA\in[\lambdaAS-\epsilon_n, \lambdaAS+\epsilon_n]}\log|\lambdaA|.\notag
\end{align}
That is, even if we assume the most conservative plant dynamics within
(\ref{uncertainty}), the bound for the known plants case is looser than the
necessary data rate bound $\Rnec$.

Moreover, we remark that for the uncertain plants such that $\Rnec<1$,
we have that $\Rnec<\max_{\lambdaA\in[\lambdaAS-\epsilon_n, \lambdaAS+\epsilon_n]}\log|\lambdaA|$.
However, we can not construct a quantizer achieving $R<1$ since this implies
that the number of the quantization cells $N$ is less than $2$.
Thus, the case of $\Rnec<1$ may be less interesting from a practical viewpoint.

In the following example, we compare the necessary limitation $\Rnec$ with
the result for known plants to confirm the gap between the two bounds.
\begin{exmp}\label{ex,neccomparison}
 Consider a plant with $\epsilon_n=0.35$.
\begin{figure}[t]
 \centering
 \includegraphics[scale=.6]{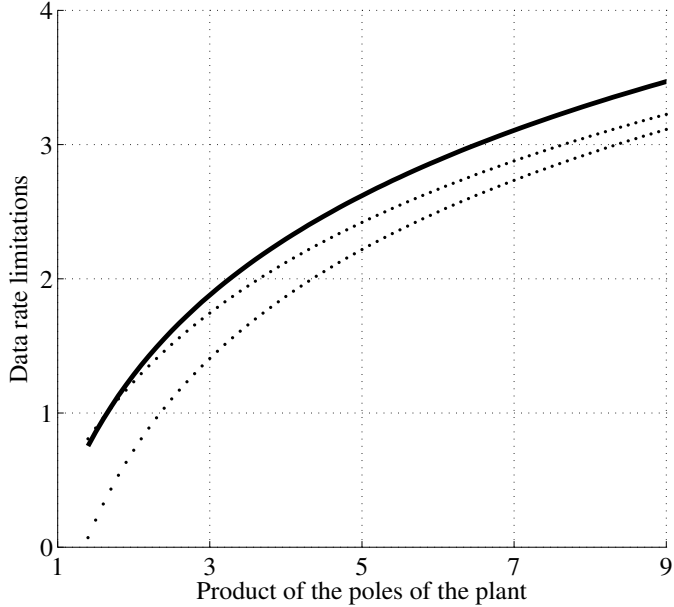}
 \caption{Data rate limitations versus the product of the poles $\lambdaAS$
 ($\epsilon_n=0.35$):
 $\Rnec$ (solid) and the maximum and the minimum of the bound $\log|\lambdaA|$
 for the known plants case within the uncertainty (dotted).}
 \label{fig,nec_comparison}
\end{figure}
\fref{fig,nec_comparison} shows the data rate bounds $\Rnec$ and $\log|\lambdaA|$
versus the product of the poles $\lambdaAS$ of the nominal plant or $a_n^*$.
The solid line represents the necessary data rate in Theorem~\ref{th,nec},
and the dotted lines are those for known plants.
Note that the bounds do not depend on the parameters $a_{1},\dots,a_{n-1}$.
Due to the uncertainty, the bound for known plants varies within the area
bounded by the dotted lines.
However, there exists a gap between the solid line and the upper dotted line.
Hence, the limitations for known plants are insufficient in the presence of
uncertainties.
Furthermore, we observe that when $R<1$ the upper dotted line is greater
than the solid line as we mentioned above.
\end{exmp}

The work \cite{Okano2012} shows another special case of Theorem~\ref{th,nec}.
for the case when the quantizer is uniform;
the uniform quantizer is the simplest
quantizer, which divides the input range into quantization cells of same lengths.
If the plant is uncertain with $\epsilon_n>0$, then the necessary data rate
bound in \cite{Okano2012} is higher than that in (\ref{nec,R}).
Therefore, we may stabilize the system with a lower data rate by using
a quantizer that is not uniform but more general.

The proof of the Theorem~\ref{th,nec} will be presented in the following
two sections.
Throughout the proof, the central question is as follows:
Under the situation where the estimation set of the plant state becomes
large due to instability, at least how precise is the quantization
required to be to make the estimation set gradually small?
To answer this question, in Section~\ref{sec,gnec}, we evaluate the expansion
rate of the estimation set for a given quantizer.
We focus on the effect of $a_n$ on the expansion of the estimation set,
since $a_n$ is equal to the product of the poles of the plant.
Then, in Section \ref{sec,opt_q}, the optimal quantizer which minimizes
the expansion rate is presented.

\section{Upper Bound on Expansion Rate for a Given Quantizer}\label{sec,gnec}
In this section, we analyze the expansion rate of the state estimation set
for a given quantizer whose boundary points are $\{h_l\}_l$.
We first introduce the sequence $w_l$, $l=0,1,\dots,\lceil N/2\rceil-1$, as
\begin{align}
  w_{l}&:=
 \begin{cases}
  2(|a_n^*|+\epsilon_n)h_{l+1} & \text{if }N\text{ is odd and }l=0,\\
  (|a_n^*|+\epsilon_n)h_{l+1}-(|a_n^*|-\epsilon_n)h_l  &  \text{else}.
 \end{cases}\label{def,w}
\end{align}
Then, the next lemma holds as a necessary condition for the quantizer
$\{h_l\}_l$.

\begin{lem}\label{lem,generalnec}
 Under Assumption~\ref{asm,instability}, if the system in \fref{fig,system}
is stable, then it holds that
\begin{align}
 \max_{l\in\{0,1,\dots,\lceil N/2\rceil-1\}}w_l<1.\label{lem1,cond}
\end{align}
\end{lem}

\begin{pf}
First, we show that stability of the system requires convergence of $\sigma_k$.
For the estimation set $\calY_k$, it is obvious that
$ \max_{y_k'\in\calY_k}|y_k'|\geq\mu(\calY_k)/2$.
Letting $\delta$ be the smallest width of the quantization cells, we have that
$\mu(\calY_k)\geq \delta\sigma_k$.
Hence, if $\lim_{k\to\infty}|y_k|=0$, then $\lim_{k\to\infty}\sigma_k=0$
holds.

In the rest of the proof, we show that (\ref{lem1,cond}) is a necessary condition
for the convergence of $\sigma_k$.
Notice that from (\ref{sigma_ineq}) $\sigma_{k+1}$ is bounded from below by
$\mu(\calY_{k+1}^-)$. 
Substitution of (\ref{BM}) from Lemma~\ref{lem,BM} into (\ref{sigma_ineq}) yields
\begin{align}
 \sigma_{k+1}\geq\sum^{n}_{i=1}\mu\left(\calA_i\calY_{k-i+1}\right)
 \geq\mu\left(\calA_n\calY_{k-n+1}\right).\label{nec,1}
\end{align}

We next evaluate the far right-hand side of the above inequality.
The set $\calA_n\calY_{k-n+1}$ depends on the boundaries of $\calA_n$ and
$\calY_{k-n+1}$.
By (\ref{uncertainty}), we have $\calA_n=[a_n^*-\epsilon_n,a_n^*+\epsilon_n]$.
The boundaries of $\calY_{k-n+1}$ vary depending the cell which the output
$y_{k-n+1}$ fell in.
Let define the index $l_{k-n+1}$ of the cell as follows:
For given $\calY_{k}$ and $\sigma_k$, let $l_k$ be the index of $\calY_k$
such that $\inf_{y'_k\in\calY_k}\left|y'_k/\sigma_k\right|=h_{l_k}$.
We claim that the width $\mu(\calA_n\calY_{k-n+1})$ can be written as
\begin{align}
 \mu(\calA_n\calY_{k-n+1})=w_{l_{k-n+1}}\sigma_{k-n+1}. \label{nec,2}
\end{align}
Here, for simplicity, we assume that $a_n^*>0$.
Notice that from (\ref{a-e>1}), $\calA_n$ does not contain the origin.
By replacing $a_n^*$ with $|a_n^*|$ in the discussion, we can obtain the
relations for the case $a_n^*<0$.

To derive (\ref{nec,2}), we consider the following two cases (i) and (ii)
and use basic results of interval arithmetics \cite{Moore1966}.
Denote the infimum and the supremum of $\calY_{k-n+1}$ by $\uY_{k-n+1}$
and $\bY_{k-n+1}$, respectively.

(i) $\uY_{k-n+1}<0<\bY_{k-n+1}$: In this case, from (\ref{def,cell_odd}),
(\ref{def,cell_even}), and (\ref{def,calY}), $N$ is odd and $l_{k-n+1}=0$,
or equivalently, $\calY_{k-n+1}=[-h_1\sigma_{k-n+1},h_1\sigma_{k-n+1})$.
Thus, the width of the product of the intervals is computed as follows:
\begin{align}
 \mu(\calA_n\calY_{k-n+1})&=(a_n^*+\epsilon_n)(\bY_{k-n+1}-\uY_{k-n+1})\notag\\
 &=2(a_n^*+\epsilon_n)h_1\sigma_{k-n+1}\notag.
\end{align}
Hence, (\ref{nec,2}) holds for this case.

(ii) $\uY_{k-n+1}\geq0$ or $\bY_{k-n+1}\leq 0$:
In this case, $N$ is even or $l_{k-n+1}\neq0$.
First, suppose that $0\leq \uY_{k-n+1}$.
Noticing (\ref{a-e>1}), we have
\begin{align}
 \mu(\calA_n\calY_{k-n+1})
 &=(a_n^*+\epsilon_n)\bY_{k-n+1}-(a_n^*-\epsilon_n)\uY_{k-n+1}\notag\\
 &=\left\{(a_n^*+\epsilon_n)h_{l_{k-n+1}+1}-(a_n^*-\epsilon_n)h_{l_{k-n+1}}\right\}\sigma_{k-n+1}.
 \notag
\end{align}
The same equalities can be established for the case of $\bY_{k-n+1}\leq 0$
by flipping the signs of $\bY_{k-n+1}$ and $\uY_{k-n+1}$.
Thus, we have (\ref{nec,2}) for this case also.

Finally, by (\ref{nec,1}) and (\ref{nec,2}), it holds that
$\sigma_{k+1}\geq w_{l_{k-n+1}}\sigma_{k-n+1}$.
Since $\sigma_k\to0$ for all possible parameters in (\ref{uncertainty})
and initial values, the maximum of $w_{l_{k-n+1}}$ must be less than $1$,
i.e., (\ref{lem1,cond}) is necessary.
\end{pf}

\section{Optimal Quantizer}\label{sec,opt_q}
In the previous section, we have seen a condition for stability on the
expansion rate $w_l$, which is defined depending on the quantizer.
In this section, we find the quantizer that minimizes $\max_l w_l$ for a fixed $N$.
To state such an optimal quantizer, we introduce the quantizer $q^*_N$
represented by the boundary points $\{h_l^*\}_l$ as follows:

(i) If $\epsilon_n>0$, then
\begin{align}
 h_{l}^*=
 \begin{cases}
  \frac{1}{2}\frac{1-tr^l}{1-tr^{\lceil N/2\rceil}} &\text{if }N \text{ is odd},\\
  \frac{1}{2}\frac{1-r^l}{1-r^{\lceil N/2\rceil}} &\text{if }N \text{ is even},
 \end{cases}\label{q*_optimal}
\end{align}
where $t:=|\lambdaAS|/(|\lambdaAS|-\epsilon_n)$.

(ii) If $\epsilon_n=0$, then
\begin{align}
 h_{l}^*=
 \begin{cases}
  \frac{1}{N}\left(l-\frac{1}{2}\right) &\text{if }N \text{ is odd},\\
  \frac{1}{N}l &\text{if }N \text{ is even}.
 \end{cases}\label{q*_uniform}
\end{align}

The following lemma holds.
\begin{lem}\label{lem,opt_q}
 The quantizer $q^*_N$ minimizes $\max_lw_l$.
\end{lem}

\fref{fig,q^*} illustrates the quantization boundaries $\{h^*_l\}_l$ of
the optimal $q^*_N$ when $|\lambdaAS|=3.0$, $\epsilon_n=0.5$, and $N=8$.
\begin{figure}[tbp]
 \centering
 \includegraphics[scale=.6]{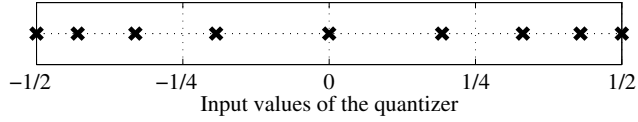}
 \caption{Boundaries of the quantizer $q^*_N$ when $|\lambdaAS|=3.0$,
 $\epsilon_n=0.5$, and $N=8$.}
\label{fig,q^*}
\end{figure}
We observe that the quantizer takes its quantization cells smaller towards
the boundaries $\pm1/2$ of the input range.
This nonuniformity is an outcome of the minimization of $\max_l w_l$.
Intuitively, this characteristic can be explained as follows.
For simplicity, consider the case of a scalar plant where the parameter
is given as $a\in\calA=[a^*-\epsilon,a^*+\epsilon]$, $a^*>0$.
Under the control scheme, the plant output $y_k$ is quantized and only the cell,
or the interval $\calY_{k}$, to which it belongs is known to the controller.
After one time step, because of the plant instability, the interval in which
the output should be included will expand in width.
When the plant model is known, the expansion ratio is constant and is equal
to $|a^*|$ for any cell.
However, with plant uncertainties, the ratio depends on the location
of the cell.
In particular, cells further away from the origin expands more.
This fact is illustrated in \fref{fig,AYuni} when the quantization is uniform.
In contrast, when the proposed quantizer $q^*_N$ is used, the intervals
after one step have the same width (see \fref{fig,AYopt}).
\begin{figure}[t]
 \begin{subfigure}{\linewidth}
  \centering
  \includegraphics[scale=1]{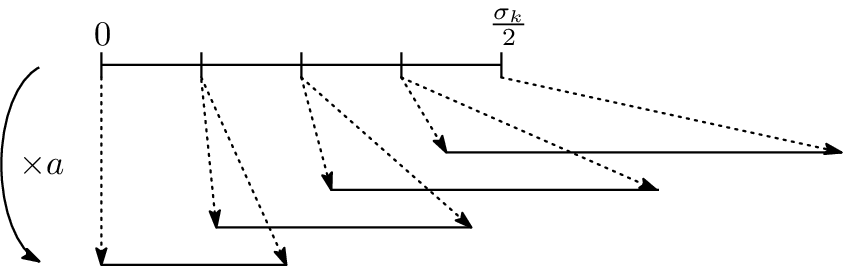}
  \caption{Uniform quantizer case.} \label{fig,AYuni}
 \end{subfigure}
 \begin{subfigure}{\linewidth}
  \centering
  \includegraphics[scale=1]{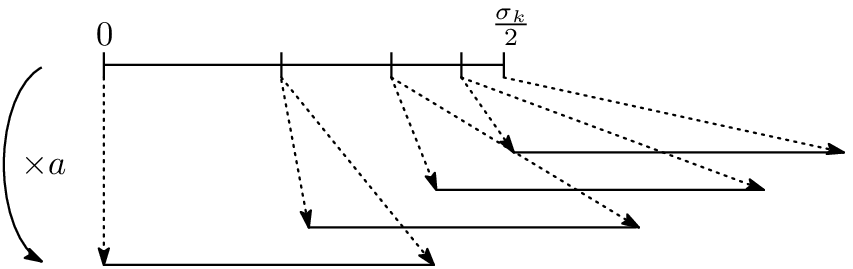}
  \caption{Optimal quantizer case.} \label{fig,AYopt}
 \end{subfigure}
 \caption{Expansion of the intervals in which the output should be included 
 by the plant instability $a\in\calA=[a^*-\epsilon,a^*+\epsilon]$, $a^*>0$.}
 \label{fig,AY}
\end{figure}
 
Furthermore, the quantizer $q^*_N$ becomes more nonuniform in the presence
of more uncertainties in the plant, expressed with a larger $\epsilon$.
This can be seen in the definition of the boundary points $\{h_l^*\}_l$,
where the ratio $r$ given in (\ref{def,nu,r}) determines
the widths of the cells.
Note that when $\epsilon=0$, $q^*_N$ becomes a uniform one as we have
seen in (\ref{q*_uniform}).

For the general order plants case, the parameter $a$ in the above explanation
should be replaced with the $n$th parameter $a_{n}$, which is equal to the product
of the poles of the plant and takes the nominal value as $|\lambdaAS|$.
Thus, the proposed quantizer $q^*_N$ minimizes the maximum width of the
intervals expanded by $a_{n}$ and other parameters do not affect the structure
of $q^*_N$.
This is because we have focused on the effect of $a_{n}$ in the stability
analysis in the proof of Lemma~\ref{lem,generalnec}.
As a consequence, $q^*_N$ is expressed in a simple form by $|\lambdaAS|$
and its uncertainty $\epsilon_n$.

Finally, it is interesting to note that as we see in \fref{fig,q^*}, the quantizer
$q^*_N$ has a property in contrast with the logarithmic quantizer studied
in \cite{Elia2001, Fu2005, Hayakawa2009, Tsumura2009};
in such quantizers, the quantization cells become small for inputs around the origin
and grow exponentially as the input size increases.
We also note that other nonuniform quantizers have been studied in \cite{Li2004}
and \cite{Tsumura2009a}, where stabilization of a continuous-time system
and a system identification problem over digital communication channels
have been respectively investigated.

\begin{pf}[Proof of Lemma~\ref{lem,opt_q}]
After some calculation, we have that $w_l$ is constant with respect to $l$,
i.e.,
\begin{align}
 w_l=w_{l'}\text{ for any } l,l'\in\{0,1,\dots,\lceil N/2\rceil-1\}\label{wconst}
\end{align}
if and only if the quantizer is $q^*_N$.
Therefore, it is enough to show that a quantizer which does not satisfy
(\ref{wconst}) yields a larger $\max_lw_l$ compared with the case $q^*_N$.
We prove this by contradiction.

For a given quantizer with the boundaries $\{h_l\}_l$,
denote the expansion rates $w_l$ under the quantizer by $w_l(h)$.
Assume that there exists a quantizer $\{g_l\}_l$ such that (\ref{wconst}) is
not satisfied and it holds that $ \max_{l}w_{l}(g)<\max_{l}w_{l}(h^*)$.
Then, for any $l\in\{0,1,\dots,\lceil N/2\rceil-1\}$,
\begin{align}
 w_{l}(g)\leq \max_{l'}w_{l'}(g)<\max_{l'}w_{l'}(h^*)=w_{l}(h^*).\label{gh^*}
\end{align}

We now look at the relation between $g_{l}$ and $h_{l}^*$ for each $l$.
From (\ref{quantizeredge}), we have $g_0=h^*_0=0$.
Substituting these equations into (\ref{def,w}), we obtain
\begin{align}
 w_0(g)&=\begin{cases}
 2(|a_n^*|+\epsilon_n)g_1 & \text{if }N\text{ is odd},\\
 (|a_n^*|+\epsilon_n)g_1  & \text{else},
\end{cases}\notag\\
 w_0(h^*)&=\begin{cases}
 2(|a_n^*|+\epsilon_n)h^*_1 & \text{if }N\text{ is odd},\\
 (|a_n^*|+\epsilon_n)h^*_1  & \text{else}.
\end{cases}\notag
\end{align}
For the case $l=0$ in (\ref{gh^*}), we have $w_0(g)<w_0(h^*)$.
Thus, from the above equations, we have that
\begin{align}
 g_1<h_1^*.\label{l=1}
\end{align}
Furthermore, by (\ref{def,w}), it follows that
\begin{align}
 w_{l}(g)&=(|a_n^*|+\epsilon_n)g_{l+1}-(|a_n^*|-\epsilon_n)g_{l},\notag\\
 w_{l}(h^*)&=(|a_n^*|+\epsilon_n)h^*_{l+1}-(|a_n^*|-\epsilon_n)h^*_{l}\notag
\end{align}
for $l\in\{1,2,\dots,\lceil N/2\rceil-1\}$.
Substitution of these equations into (\ref{gh^*}) gives
\begin{align}
 g_{l+1}\leq rg_{l}+\frac{\max_{l'}w_{l'}(g)}{|a_n^*|+\epsilon_n},\
 h^*_{l+1}=rh^*_{l}+\frac{\max_{l'}w_{l'}(h^*)}{|a_n^*|+\epsilon_n}.\notag
\end{align}
By introducing the relation (\ref{l=1}) to the above, we recursively obtain
$g_{l}<h^*_{l}$ for all $l\in\{1,2,\dots,\lceil N/2\rceil\}$.
This contradicts $g_{\lceil N/2\rceil}=h^*_{\lceil N/2\rceil}=1/2$
given in (\ref{quantizeredge}). Therefore, it follows that $\{h_l^*\}_l$
is the optimal quantizer.
\end{pf}

Since we found the quantizer which minimizes $\max_l w_l$, the lower bound
on $N$ satisfying (\ref{lem1,cond}) is the necessary condition on the data
rate $R$ $(=\log N)$.
This is to be proved as the last step of the proof of Theorem~\ref{th,nec}.

\begin{pf}[Proof of Theorem~\ref{th,nec}]
 In this proof, we derive the bounds (\ref{nec,R}) and (\ref{nec,e}) from
(\ref{lem1,cond}) in Lemma \ref{lem,generalnec} by employing $q^*_N$ as
the quantizer.
First, suppose that $\epsilon_n>0$.
We consider the following two cases.

(i) $N$ is even: In this case, by the definition of $\{h_l^*\}_l$, we have
that $h_l^*=(1-r^l)/(1-r^{\lceil N/2\rceil})$. Thus, it holds that
$\max_lw_l={\epsilon_n}/(1-r^{\lceil N/2\rceil})$.
Consequently, the necessary condition (\ref{lem1,cond}) is equivalent to 
\begin{align}
 N>\Nnece:=\frac{\log(1-\epsilon_n)^2}{\log r},\ \epsilon_n<1.\notag
\end{align}

(ii) $N$ is odd:
Similarly, we have that the inequalities
\begin{align}
 N>\Nneco:=\frac{\log\left\{(1-\epsilon_n)/t\right\}^2}{\log r},\ \epsilon_n<1\notag
\end{align}
are necessary.

Comparing $\Nnece$ with $\Nneco$, it is clear that $\Nneco>\Nnece$.
Hence, $N>\Nnece$ is necessary for both cases (i) and (ii).
From the relation $R=\log N$, we obtain the condition (\ref{nec,R}).

For the case $\epsilon_n=0$, noticing that $\max_lw_l={|a_n^*|}/{N}$,
we have (\ref{nec,R}) for this case also.
\end{pf}

\section{Stabilizing Controller: Fixed Data Rate Case}\label{sec,suf}
In this section, we present a sufficient condition for the existence of
a stabilizing feedback control scheme under the static data rate.
The condition offers a stability test by calculating the spectral radius
of a certain matrix.

Given a certain data rate $R$, or $N$, and a static quantizer $\{h_l\}_l$,
we employ the control law as follows: In the encoder (\ref{def,s}), the
scaling parameter is determined by
\begin{align}
 \sigma_k=&\mu(\calY^-_{k}),\label{suf,sigma}
\end{align}
and in the controller (\ref{controller}), the control is given as
\begin{align}
 u_k=&-\frac{1}{2}\left(\bY^-_{k+1}+\uY^-_{k+1}\right).\label{suf,u}
\end{align}
Here, we denote the supremum and the infimum of $\calY_{k+1}^-$ as
$\bY^-_{k+1}$ and $\uY^-_{k+1}$, respectively.

Next, we introduce some notations required for the analysis of the resulting system.
For $i=1,2,\dots,n$, define $w_{i,l}$, $l=0,1,\dots,\lceil N/2 \rceil-1$,
as follows:
If $N$ is odd, $w_{i,0}:=2(|a_i^*|+\epsilon_i)h_{1}$, and for $l\geq1$,
\begin{align}
 w_{i,l}:=
 \begin{cases}
  (|a_i^*|+\epsilon_i)h_{l+1}-(|a_i^*|-\epsilon_i)h_{l} & \textnormal{if } \calA_i\not\ni 0,\\
  2\epsilon_ih_{l+1} & \textnormal{if } \calA_i\ni 0.
 \end{cases}\label{def,wil}
\end{align}
If $N$ is even, $w_{i,l}$ is defined as in (\ref{def,wil}) for any $l$.
We note that $w_{l}$ defined in (\ref{def,w}) corresponds to $w_{n,l}$.
As we will see in later, $w_{i,l}$ represents the expansion rate of the
quantization cells, $[h_{l},h_{l+1})$ and $[-h_{l+1},-h_l)$, enlarged
by the parameter $a_i$.
We denote by $\widebar w_i$ the maximum of $w_{i,l}$ over all $l$.
Moreover, define the matrix $H\in\R^{n\times n}$ containing $\widebar w_1,\dots,\widebar w_n$ as
\begin{gather}
 H:=\left[\begin{array}{cccc}
   0& 1& \cdots& 0\\
   \vdots& \ddots& \ddots& \vdots\\
   0& 0& \cdots& 1\\
   \widebar w_n & \widebar w_{n-1} & \cdots& \widebar w_1
 \end{array}\right].\label{def,H}
\end{gather}

We are now ready to present the main theorem of this section.
\begin{thm}\label{th,suf}
 Given the data rate $R=\log N$ and the quantizer $\{h_l\}_l$,
if the matrix $H$ in (\ref{def,H}) satisfies
\begin{align}
 \rho(H)<1\label{suf_cond},
\end{align}
then under the control law using (\ref{suf,sigma}) and (\ref{suf,u}),
the system depicted in \fref{fig,system} is stable.
\end{thm}

We emphasize that the optimal quantizer $q_N^*$ proposed in Lemma~\ref{lem,opt_q}
can reduce the sufficient data rate compared with the uniform one,
employed in \cite{Martins2006, Phat2004, Okano2012}.
To confirm this, we now show an example.

\begin{exmp}\label{ex,necsuf}
Consider a second-order plant, where the uncertainty bounds are taken as
$\epsilon_1=0.10$ and $\epsilon_2=0.35$.
\begin{figure}[t]
 \centering
 \includegraphics[scale=.6]{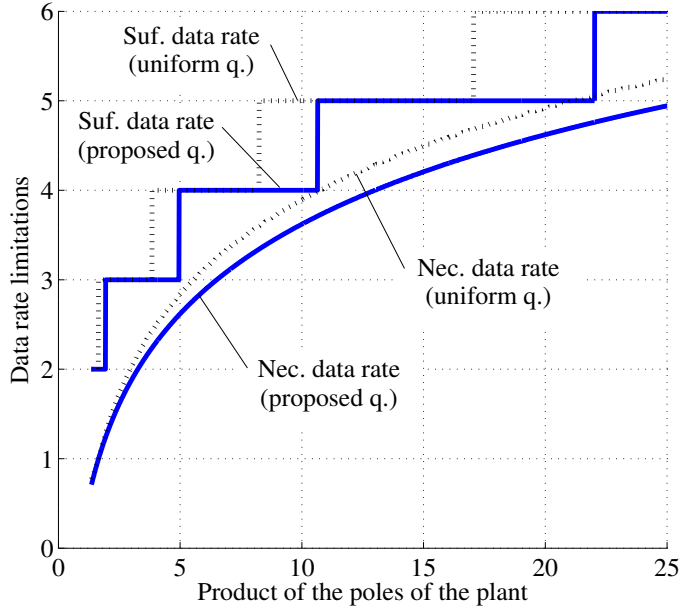}
 \caption{Bounds on the data rate ($n=2$, $a_1^*=1.0$, $\epsilon_1=0.10$,
 $\epsilon_2=0.35$):
 The solid lines are the sufficient bound and the necessary bound when
 the quantizer is optimal $q_N^*$, while the dotted lines are those for
 the case of the uniform one.}
 \label{fig,necSufDatarateOptUni}
\end{figure}
We fix $a_1^*$ as $1.0$ and plot the bounds on the data rate
versus the product of the poles of the nominal plant $|\lambdaAS|=|a_2^*|$
in \fref{fig,necSufDatarateOptUni}.
The sufficient bounds have been obtained by computing the minimum $R$ satisfying
(\ref{suf_cond}) numerically.
In the figure, the solid lines illustrate the bounds given by Theorems \ref{th,nec}
and \ref{th,suf} when the quantizer is $q_N^*$, and the dotted lines are
those for the uniform case studied in \cite{Okano2012}.
The figure shows that by using the optimal quantizer, we can stabilize
the system under a lower data rate compared with the case using the uniform
one.
Note that the sufficient bounds take discrete values since the rates are
rounded to integers.
In Section \ref{sec,averagesuf}, we discuss the gap due to this integer constraint.
\end{exmp}

\begin{pf}[Proof of Theorem~\ref{th,suf}]
First, we show that if $\sigma_k\to 0$ then $y_k\to0$ as $k\to\infty$
under the control law.
This is easy to establish because by substituting (\ref{suf,u}) to (\ref{AR})
and by referring to the definition (\ref{def,calY^-}) of $\calY_{k+1}^-$,
we have that
\begin{align}
 |y_{k+1}|
 &=\left|a_{1}y_k+\cdots+a_{n}y_{k-n+1}-\frac{1}{2}\left(\bY^-_{k+1}+\uY^-_{k+1}\right)\right|\notag\\
 &\leq\frac{\mu(\calY_{k+1}^-)}{2}=\frac{\sigma_{k+1}}{2}.\notag
\end{align}

Next, we prove that (\ref{suf_cond}) implies that $\sigma_k\to0$.
By (\ref{suf,sigma}) and the equality in (\ref{BM}) from Lemma~\ref{lem,BM},
we have
\begin{align}
 \sigma_{k+1}=\sum^{n}_{i=1}\mu\left(\calA_i\calY_{k-i+1}\right).\label{suf,sigmamu}
\end{align}
For the $i$th term $\mu\left(\calA_i\calY_{k-i+1}\right)$, we claim that
\begin{align}
 \mu\left(\calA_i\calY_{k-i+1}\right)=w_{i,l}\sigma_{k-i+1}.\label{muAY_wilsigma}
\end{align}
Here, $l$ is the index of the boundary of $\calY_{k-i+1}$ closer
to the origin, i.e., $l$ is the index such that
$h_{l}\sigma_{k-i+1}=\min\{|\widebar\calY_{k-i+1}|,\,|\wideubar\calY_{k-i+1}|\}$.

To show (\ref{muAY_wilsigma}), we first compute $\calA_i\calY_{k-i+1}$
using basic results in interval arithmetics \cite{Moore1966}.
Then we have
\begin{align}
 \mu\left(\calA_i\calY_{k-i+1}\right)
 =\begin{cases}
    \left(|a_i^*|+\epsilon_i\right)\mu(\calY_{k-i+1})
    & \text{if }\wideubar\calY_{k-i+1}<0<\widebar\calY_{k-i+1},\\
    |a_i^*|\mu(\calY_{k-i+1})+\epsilon_i|\widebar\calY_{k-i+1}+\wideubar\calY_{k-i+1}|
    &  \text{else if }\calA_{i}\not\ni 0,\\
    2\epsilon_i\max\left\{|\widebar\calY_{k-i+1}|,\,|\wideubar\calY_{k-i+1}|\right\}
    &  \text{else}.
   \end{cases}\label{muAY}
\end{align}
Furthermore, we consider the following two cases and show (\ref{muAY_wilsigma})
from (\ref{muAY}) for each case.

(i) $\wideubar\calY_{k-i+1}<0<\widebar\calY_{k-i+1}$:
From the symmetry of the quantizer, $N$ must be odd and $l=0$,
and hence, $\mu(\calY_{k-i+1})=2h_1\sigma_{k-i+1}$.
By substituting this equality to the first case in (\ref{muAY}),
(\ref{muAY_wilsigma}) is established.

(ii) Otherwise:
The absolute values of the boundaries of $\calY_{k-i+1}$
are $h_{l}\sigma_{k-i+1}$ and $h_{l+1}\sigma_{k-i+1}$.
Noticing that $0\leq h_{l}<h_{l+1}$, we have (\ref{muAY_wilsigma})
from (\ref{muAY}) for this case also.

By taking the maximum of the right-hand side of (\ref{muAY_wilsigma}) over
all possible $y_{k-i+1}$, or all $l$,
and by (\ref{suf,sigmamu}), we have an upper bound of $\sigma_{k+1}$ as
\begin{align}
 \sigma_{k+1}\leq\sum^{n}_{i=1}\widebar w_{i}\sigma_{k-i+1}.\label{sigma-theta}
\end{align}
Here, consider the following linear system
\begin{align}
 \zeta_{k+1}=H\zeta_k,\quad
 \zeta_0=\left[\sigma_{-n+1}\ \sigma_{-n}\ \cdots\ \sigma_{0}\right]^T.
 \notag
\end{align}
From (\ref{sigma-theta}), the bottom element of $\zeta_{k+1}$ is greater
than or equal to $\sigma_{k+1}$ for any time $k$.
The inequality (\ref{suf_cond}) implies that $H^k$ converges to the zero
matrix and hence $\sigma_k\to0$ as $k\to\infty$.
\end{pf}

\section{Tightness of the Limitations}\label{sec,scalar}
Here, we discuss tightness of the derived limitations.
We first present the special cases that Theorems \ref{th,nec} and
\ref{th,suf} become necessary and sufficient conditions.
We then compare tightness of the limitations with that of the existing
works dealing with plant uncertainty.

As we have seen in Example \ref{ex,necsuf}, in general, there exists a gap between
the necessary bound given by Theorem~\ref{th,nec} and the
sufficient bound obtained from Theorem~\ref{th,suf}.
This gap is caused since, in both necessity and sufficiency analyses,
there exist conservativeness in the evaluation of the width of the prediction
set $\calY^-_{k+1}$.

Theorem~\ref{th,suf} becomes a tight condition for stability in the following
case:
For the class of plants and quantizers where the indices of the quantization
cells which result in the maximum expansion rates, i.e., $\arg\max_l w_{i,l}$,
are the same for all $i=1,2,\dots,n$, the condition (\ref{suf_cond}) is
also necessary for stability.
This fact is followed by the proof of Theorem \ref{th,suf} and
the discussion below.
In the proof, we evaluate the width $\mu(\calY^-_{k+1})=\sum_{i=1}^n\mu(\calA_i\calY_{k-i+1})$
over all possible $y_{k-n+1},\dots,y_n$.
As an upper bound on the width, we consider the sum of the
maximum $\widebar w_{i}\sigma_{k-i+1}$ of each summand $\mu(\calA_i\calY_{k-i+1})$.
When $\arg\max_l w_{i,l}$ are the same for all $i$, this upper bound
and hence the condition (\ref{suf_cond}) become tight.

We next present the special case that the limitations given in Theorems
\ref{th,nec} and \ref{th,suf} coincide.
In the derivation of the necessary result Theorem \ref{th,nec},
we have evaluated a lower bound on $\mu(\calY^-_{k+1})$ by focusing only on
the $n$th parameter $a_{n}$.
This approximation leads us to the explicit limitations but causes conservativeness.
In light of this point, if
\begin{align}
 a_1^*=\cdots=a_{n-1}^*=0,\quad \epsilon_1=\cdots=\epsilon_{n-1}=0,\label{escalarsys}
\end{align}
then the limitations (\ref{nec,R}) and (\ref{nec,e}) in Theorem \ref{th,nec}
are equivalent to the inequality (\ref{suf_cond}) in Theorem \ref{th,suf} and tight.

This fact is stated as a corollary below.
\begin{cor}\label{cor,scalar}
 Under Assumption \ref{asm,instability}, in the system depicted in \fref{fig,system},
if the plant satisfies (\ref{escalarsys}), then the following hold:
\begin{itemize}
 \item[(i)]  The system is stable if and only if  $R>\lceil \Rnec\rceil$
	     and (\ref{nec,e}) hold.
 \item[(ii)] The quantizer $q^*$ minimizes the required data rate for
stability.
\end{itemize}
\end{cor}

Note that the plants satisfying (\ref{escalarsys}) can be reduced to the scalar
plants ($n=1$).
Hence, it is enough to show that Theorem \ref{th,nec} becomes tight
for the scalar plants case.

\begin{pf}
Consider the case $n=1$.
(i) We prove the sufficiency.
From the proof of Theorem~\ref{th,nec}, if the inequalities $R>\lceil \Rnec\rceil$
and (\ref{nec,e}) hold, then under the quantizer $q^*_{2^R}$ given in (\ref{q*_optimal}),
we have that $\max_l w_l<1$, i.e., (\ref{lem1,cond}) in Lemma~\ref{lem,generalnec}
holds.
On the other hand, since $n=1$, it follows that $\rho(H)=\widebar w_1=\max_lw_l$.
Thus, the inequality (\ref{lem1,cond}) is equivalent to the sufficient condition
(\ref{suf_cond}).

(ii) This is obvious from the fact that (\ref{suf_cond}) is equivalent
to (\ref{lem1,cond}), and Lemma~\ref{lem,opt_q}.
\end{pf}

The works of \cite{Phat2004} and \cite{Martins2006} have shown sufficient
conditions for stabilization of uncertain plants via finite data rate
channels.
We remark that those conditions contain conservativeness even for the
scalar plants case.
For the case $n=1$, the sufficient bound in \cite{Phat2004} is
\begin{align}
 R_{\text{suf}}:=\log\frac{|\lambdaAS|-\epsilon_1(|\lambdaAS|+\epsilon_1)}
 {1-\epsilon_1(2|\lambdaAS|+2\epsilon_1+1)},\notag%
\end{align}
and the one from \cite{Martins2006} becomes
\begin{align}
 R_{\text{suf}}':=\log\frac{|\lambdaAS|}{1-\epsilon_1}.\notag%
\end{align}
On the other hand, from Corollary~\ref{cor,scalar}, we have that $\Rnec$ is
a sufficient data rate bound for the case $n=1$.
It is easy to verify that $\Rnec<R_{\text{suf}}$ and $\Rnec<R_{\text{suf}}'$.
Hence, our result is tighter than the known bounds $R_{\text{suf}}$ and $R_{\text{suf}}'$.
It should be noted that, for general order plants, it is difficult to compare
Theorem~\ref{th,suf} with the bounds in \cite{Phat2004} and \cite{Martins2006}
since the types of uncertainties are different:
In \cite{Phat2004}, unstructured uncertainties have been considered, and it is
hard to describe the data rate limitation in an explicit form,
while in \cite{Martins2006}, scalar nonlinear plants have been studied and
the multi-dimensional case has not been investigated.

\section{Minimum Data Rate: Variable Data Rate Case}\label{sec,average}
So far, we have considered the case that the data rate is static.
In this and the following section, we generalize the problem setup to allow
us to employ variable data rates and then study limitations on the average
data rate for stability.

When we consider a practical quantization and communication scheme, we
can not choose the bit-length of data expressing quantized states to be
a noninteger.
Hence, for a number $R$ satisfying the sufficient condition in Theorem
\ref{th,suf}, the actual data rate required for stabilization becomes $\lceil R\rceil$,
which may be larger than $R$ as we have seen in Example~\ref{ex,necsuf}.
In various data rate results such as those in \cite{Tatikonda2004, Nair2004, You2010},
it is known that when we know the exact plant model and employ variable
data rates, then this gap on the data rate can be made arbitrarily small.
That is, for any $R\in\R$ greater than the bound %
there exists a feasible controller and a pair of the encoder and the decoder
to stabilize the system.

In this section, we follow such an approach for the case of uncertain plants
and develop a control scheme with a variable data rate.
Denote by $q_{k,N_k}$ the quantizer at time $k$, where $N_k$ is the number
of the quantization cells.
In the following theorem, we show a necessary condition for stability,
which describes a bound on the average data rate
$\widebar R:=\lim_{k\to\infty}\frac{1}{k}\sum^{k-1}_{i=0}\log N_i$.

\begin{thm}\label{th,necm}
 For the system in \fref{fig,system} satisfying Assumption~\ref{asm,instability},
if the system is stable with a variable data rate, then the following inequalities
hold:
\begin{align}
 &\widebar R >\Rnec,\label{necm,R}\\
 &0\leq\epsilon_n<1,\label{necm,e}
\end{align}
where the bound $\Rnec$ is defined in (\ref{nec,R}).
\end{thm}

The above theorem implies that the necessary bounds given in Theorem~\ref{th,nec}
are valid even if we extend the communication scheme from static to time varying.
We note that in the literature there are two types of communication schemes with
variable data rates.
The first type is the one considered here and also in \cite{Tatikonda2004}.
Under the scheme, for every time step, the output is observed and the
control input is applied.
On the other hand, another communication scheme has been developed in
\cite{Nair2004, You2010, Okano2012}.
The idea is as follows:
Divide the time into cycles of a certain duration.
At the initial step of each cycle, the encoder observes the output and
then sends it slowly during the cycle.
At the end of the cycle, the controller estimates the plant state
and generates the input using the received information regarding the state
data of the state from the initial time of the cycle.
Except the last steps of the cycles, the control input is kept as zero.

It is interesting that when the plant is known, these two schemes
provide the same bound on the data rate, but for uncertain plants, the bounds
are different.
This is because in the latter scheme, the plant uncertainty causes
accumulation of error in the state estimation, which affects the accuracy
in the control input, since only the information at the beginning of the
cycle is used.

When the plant has no uncertainty, the proof of Theorem~\ref{th,necm} follows
in a straightforward manner from that of Theorem~\ref{th,nec}.
However, in the case of uncertain plants, we need a few additional steps.
As in the the static data rate case, the nonuniform quantizer $q_N^*$ in
(\ref{q*_optimal}) will be shown to be optimal in the derivation, but its
complex definition brings some analytical difficulties.

We will prove Theorem~\ref{th,necm} in two steps in the following.
As the first step, we give another form of the necessary condition as a lemma.
To state the lemma, we introduce $v_k$, which is the maximum $w_l$ in (\ref{def,w})
when the quantizer is $q_{N_k}^*$:
\begin{align}
 v_k:=\begin{cases}
  \frac{\epsilon_n}{1-tr^{(N_{k}+1)/2}}
   & \text{if } \epsilon_n>0\text{ and }N_{k}\text{ is odd},\\
  \frac{\epsilon_n}{1-r^{N_{k}/2}}
   & \text{if } \epsilon_n>0\text{ and }N_{k}\text{ is even},\\
  \frac{|a_n^*|}{N_{k}} & \text{if }\epsilon_n=0.
 \end{cases}\label{def,v}
\end{align}

Then the following lemma holds.
\begin{lem}\label{lem,necm_pre}
 For the system in \fref{fig,system} satisfying Assumption~\ref{asm,instability},
if the system is stable with variable data rates,
then for each $\alpha\in\{0,1,\dots,n-1\}$, there exists a sequence of integers
$\{m^{(\alpha)}_{i}\}_{i=0}^\infty$ such that
\begin{align}
 \prod_{j=0}^{m^{(\alpha)}_{i}-1} v_{n(\sum_{l=0}^{i-1}m^{(\alpha)}_{l}+j)+\alpha}<1,\
 i=0,1,\dots,
 \label{necm_prod}
\end{align}
where $\sum_{l=0}^{-1}m^{(\alpha)}_{l}:=0$ for all $\alpha$.
\end{lem}

This lemma is a modified version of Lemma~\ref{lem,generalnec} as the
data rate is time varying and the quantizer is optimal, i.e., $q_{k,N_k}=q^*_{N_k}$.
As we have seen in the proof of Theorem~\ref{th,nec}, we focus on the effect
of the $n$th parameter $a_{n}$ on the expansion of the prediction set $\calY_{k+1}^-$.
If the quantizer is static, then the expansion rate
$v_{k-n+1}$ of $\calY_{k+1}^-$ must be smaller than $1$ for every time step.
On the other hand, when we employ a time-varying one, the expansion rate
may be greater than $1$ at a certain time.
However, for any time step there exists a time interval such that
the expansion rate from the time to the last step of the interval
becomes smaller than $1$.
\begin{figure}[t]
 \centering
 \includegraphics[scale=1]{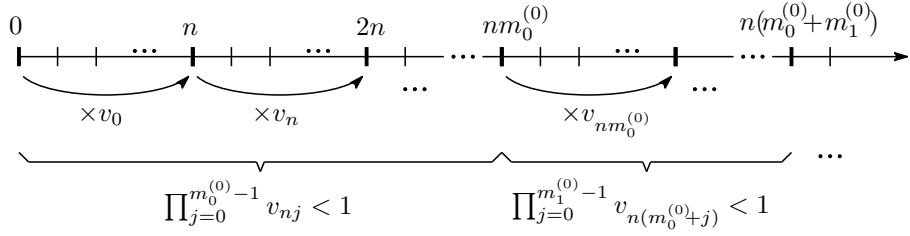}
 \caption{Intervals such that the expansion rates from the initial
 to the last steps are smaller than 1.}
 \label{fig,t-v_interval}
\end{figure}
Such an interval starting from $k=0$ is illustrate in \fref{fig,t-v_interval}.
We denote the last step of the interval by $nm^{(0)}_{0}$, where $m^{(0)}_0$ is a
positive integer.
If we take $nm^{(0)}_0$ as the initial time, we have another interval.
Let the length of the interval be $nm^{(0)}_1$.
Repeating the process we can divide the time into the intervals, where
the $i$th interval starts at $n\sum_{l=0}^{i-1}m^{(\alpha)}_{l}$ and
its length is $nm^{(0)}_i$.
Furthermore, since the index of the expansion rate $v_k$ is taken to be
$n$ periodic, there exist $n$ series of such intervals depending on the initial
time $\alpha\in\{0,1,\dots,n-1\}$ of the first interval of $i=0$
(see \fref{fig,t-v_intervals_alpha}).

\begin{figure}[t]
 \centering
 \includegraphics[scale=1]{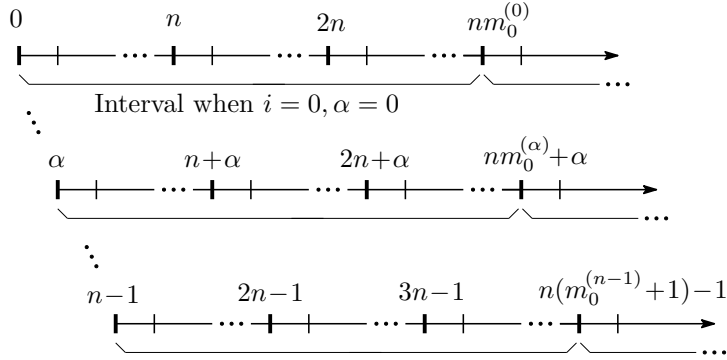}
 \caption{Intervals indexed by the initial time $\alpha$.}
 \label{fig,t-v_intervals_alpha}
\end{figure}

\begin{pf}[Proof of Lemma~\ref{lem,necm_pre}]
By the proofs of Lemma \ref{lem,generalnec} and Theorem \ref{th,nec}, we have that
\begin{align}
 \sigma_{k+1} \geq v_{k-n}\sigma_{k-n} \geq \cdots \geq
 \prod_{j=0}^{\lfloor k/n\rfloor}v_{nj+[k]_n}\sigma_{[k]_n},\notag
\end{align}
where $[\cdot]_n$ is the residue modulo $n$ and $\lfloor\cdot\rfloor$
is the floor function.
Note that $\sigma_{k+1}\to0$ as $k\to\infty$ because of the stability of the system,
but $\sigma_{[k]_n}$ remains positive since $\sigma_k$ satisfies (\ref{sigma_ineq})
and the lengths of the initial state estimation sets are positive.
Taking the limits of both sides as $k\to\infty$, we have 
\begin{align}
 \lim_{k\to\infty}\prod_{j=0}^{\lfloor k/n\rfloor}v_{nj+[k]_n}=0.\notag%
\end{align}
Hence, for each $[k]_n=\alpha\in\{0,1,\dots,n-1\}$, there exists an integer
$m^{(\alpha)}_{0}$ such that
\begin{align}
 \prod_{j=0}^{m^{(\alpha)}_{0}-1}v_{nj+\alpha}<1.\notag
\end{align}

Now, taking $m^{(\alpha)}_{0}$ as the initial time and applying the same procedure,
we have that (\ref{necm_prod}) holds.
\end{pf}

Now, we proceed to the second step to show Theorem~\ref{th,necm}.
Here we consider the cases of $\epsilon_n=0$ and $\epsilon_n>0$ separately
since the definition of the quantizer $q^*_N$, which we employ in this step,
is different depending on $\epsilon_n$.
When $\epsilon_n=0$, Theorem~\ref{th,necm} is established directly from
Lemma \ref{lem,necm_pre}:
In this case, from (\ref{def,v}), we have
$\prod_{j=0}^{m-1} v_j=|a_n^*|^{m}/2^{\sum_{j=0}^{m-1}\log N_j}$ for any
integers $m\geq0$.
Thus, the inequality (\ref{necm_prod}) in the lemma implies (\ref{necm,R}).

For the case where $\epsilon_n>0$, the upper bound (\ref{necm,e}) on
$\epsilon_n$ can be shown
by Lemma~\ref{lem,necm_pre} and the following discussion:
Taking the limit of $v_k$ as $N_k\to\infty$, we obtain
\begin{align}
 \lim_{N_k\to\infty}v_k
 \geq\lim_{N_k\to\infty}\frac{\epsilon_n}{1-r^{N_k/2}}
 =\epsilon_n.\label{lem4,e<1}%
\end{align}
The first inequality is due to the fact that $v_k\geq\epsilon_n/(1-r^{N_k/2})$
for each $N_k$.
Since $\epsilon_n/(1-r^{N_k/2})$ is monotonically decreasing with respect
to $N_k$, by (\ref{lem4,e<1}), if $\epsilon_n\geq 1$ then $v_k\geq1$ for any $N_k$.
On the other hand, if the system is stable, (\ref{necm_prod}) must hold
for a certain set of $\{N_k\}_k$ and hence, we have $\epsilon_n<1$.

From the discussion above, it is enough to establish the bound (\ref{necm,R})
on $\widebar R$ for the case $0<\epsilon_n<1$.
To do so, we evaluate the minimum of $\widebar R$ under the condition that
(\ref{necm_prod}) holds.
Here, the lower bound will be obtained by solving a certain minimization
problem.

For simplicity of notation, let $m$ be the number of elements in the interval
$m^{(\alpha)}_{i}$, and let $\hN_j$ be the number of quantization cells
$N_{n(\sum_{l=0}^{i-1}m^{(\alpha)}_{l}+j)+\alpha}$.
Then, we consider the following minimization problem:
\begin{align}
 \text{minimize}\quad &\phi(\hN):=\prod_{j=0}^{m-1} \hN_j^{1/m},\label{necm_phi}\\
 \text{subject to}\quad &\psi(\hN):=\prod_{j=0}^{m-1}\hat v_j-1\leq0.\label{necm_psi}
\end{align}
Here, we introduced the vector $\hN:=[\hN_0 \; \hN_1\; \cdots \; \hN_{m-1}]^T$ and
$\hat v_j:= {\epsilon_n}/(1-r^{\hN_j/2})$.
As a constraint, we employ (\ref{necm_psi}) with $\hat v_j$ instead of
(\ref{necm_prod}) with $v_j$.
This is because $v_j$ is difficult to treat since it depends on whether
$N_j$ is even or odd.
One can easily verify that the constraint (\ref{necm_psi}) is looser than
(\ref{necm_prod}).
Thus, the solution of the minimization problem (\ref{necm_phi}), (\ref{necm_psi})
gives a lower bound on the average data rate.

We now show that the solution $\hN^*$ is represented by using $2^{\Rnec}$.
\begin{lem}\label{lem,opt_solution}
 Consider the plant with $0<\epsilon_n<1$.
 The solution of the minimization problem (\ref{necm_phi}), (\ref{necm_psi})
is $\hN^*=[2^{\Rnec}\; 2^{\Rnec}\; \cdots \; 2^{\Rnec}]^T$.
\end{lem}

\begin{pf}
 It is obvious that $\phi(\hN)$ and $\psi(\hN)$ are convex functions.
Let $L(z,\lambda)$ be the Lagrangian of the minimization problem as
\begin{align}
 L(\hN,\lambda):=\phi(\hN)+\lambda\psi(\hN)\notag
\end{align}
and let
\begin{align}
 \hN':=\left[2^{\Rnec}\ 2^{\Rnec}\ \cdots\ 2^{\Rnec}\right]^T,\quad
 \lambda':=\frac{-\epsilon_n}{m(1-\epsilon_n)\log r}.\notag
\end{align}
Then, for $i=0,\dots,m-1$, we have that
\begin{align}
 \frac{\partial}{\partial \hN_i}L(\hN,\lambda)
 =\frac{\phi(\hN)}{m\hN_i}+\lambda\frac{\epsilon_n(\log r)r^{\hN_i/2}}{2(1-r^{\hN_i/2})^2}
 \prod_{\substack{j=0\\ j\neq i}}^{m-1}\hat v_j\notag
\end{align}
and hence $\nabla_{\!\hN}L(\hN',\lambda')=0$.
Furthermore, since $0<\epsilon_n<1$ and $\log r<0$, it holds that $\lambda'>0$.
Thus, the pair $(\hN',\lambda')$ satisfies the KKT condition \cite{Luenberger1997}
and hence, $\hN'$ is the solution.
\end{pf}

By Lemma~\ref{lem,opt_solution}, we have that $\log(\prod_{j=0}^{m-1}\hN^*_j)^{1/m}=\Rnec$
is a lower bound on the average data rate of the interval of length $m$.
Applying this result to all intervals of lengths $m^{(\alpha)}_{i}$,
we obtain (\ref{necm,R}).
This completes the proof of Theorem~\ref{th,necm}.

\section{Stabilizing Controller: Variable Data Rate Case}\label{sec,averagesuf}
In this section, we present a sufficient condition for the case of time-varying
quantization.
Here, we follow the control and communication scheme proposed in \cite{Tatikonda2004},
which is based on an $m$-periodic quantizer $\{q_{j,N_j}\}_{j=0}^{m-1}$;
at time $k$, the quantizer is $q_{[k]_m,N_{[k]_m}}$, where $[\cdot]_m$
is the residue modulo $m$.
As the scaling parameter and the control input, we employ the ones
given in (\ref{suf,sigma}) and (\ref{suf,u}) from the static quantization case.
The sufficient condition can be proved in a way similar to that in Section~\ref{sec,suf}.

We introduce slightly different notations:
For $j=0,1,\dots,m-1$, let us define $H_{j}\in\R^{n\times n}$ as
\begin{gather}
 H_{j}:=
 \left[
 \begin{array}{cccc}
   0& 1& \cdots& 0\\
   \vdots& \ddots& \ddots& \vdots\\
   0& 0& \cdots& 1\\
   \widebar w_{n,j} & \widebar w_{n-1,j} & \cdots &\widebar w_{1,j}
 \end{array}
 \right],\notag
\end{gather}
where $\widebar w_{i,j}$, $i=1,\dots,n$, are
the maximums $\widebar w_i$ of the expansion rates defined in Section \ref{sec,suf}
when the quantizer is $q_{j,N_j}$.

The following theorem holds for the variable data rate case.
\begin{thm}\label{th,suf_m}
 Given the set of quantizers $\{q_{j,N_j}\}_{j=0}^{m-1}$, if
\begin{align}
 \rho(\prod_{i=1}^{m}H_{m-i})<1\label{suf_cond_m},
\end{align}
then under the control law using (\ref{suf,sigma}) and (\ref{suf,u}),
the system depicted in \fref{fig,system} is stable.
\end{thm}

We now show a numerical example and confirm that the time-varying scheme
reduces the required data rate compared with the case of the static one.
In particular, the sufficient bound given in Theorem~\ref{th,suf_m} is
strictly lower than that given in Theorem~\ref{th,suf} except at the points
where the bounds become integers.
\begin{exmp}
Consider a scalar uncertain plant, where $\epsilon_1=0.35$.
In \fref{fig,mstep}, we plot bounds on the average data rate
versus the pole $\lambdaAS=a^*_1$ of the nominal plant.
The solid line is the achievable average data rate which is the minimum over
the duration $m\leq 10^{16}$ and quantizers $\{q_{N_j}^*\}_j$.
\begin{figure}[t]
 \centering
 \includegraphics[scale=.6]{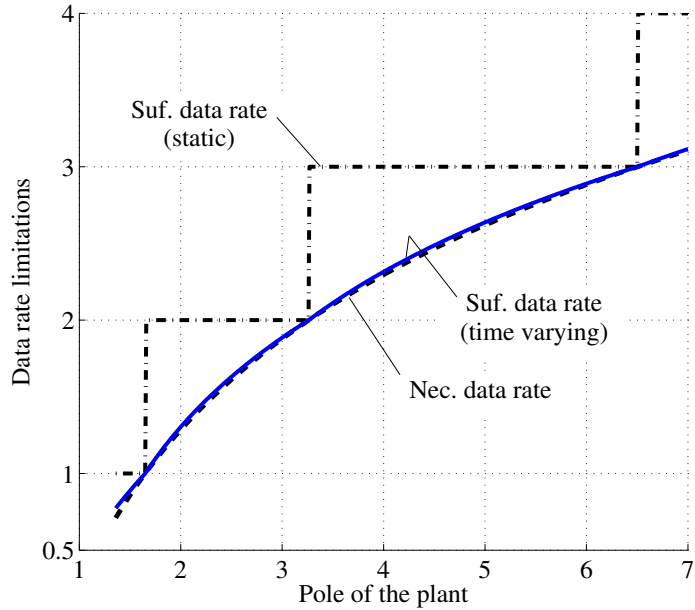}
 \caption{Limitations on the average data rate ($n=1$, $\epsilon_1=0.35$):
 The sufficient bound with the time-varying scheme (solid) is closer
 to the necessary bound (dashed) than that with static scheme (dash-dot).}
 \label{fig,mstep}
\end{figure}
The dash-dot and the dashed lines represent the sufficient bound for
the case of static scheme and the necessary bound $\Rnec$, respectively.
Note that currently we consider the case $n=1$, and hence the sufficient
bound is equal to $\lceil \Rnec\rceil$ (see Corollary~\ref{cor,scalar}).
The figure shows that the sufficient average data rate (solid line) is
smaller than that for static schemes.
Moreover, it is close to the necessary bound (dashed line) although there
exists a gap between them.
Note that, when the plant has no uncertainty, the gap can be arbitrarily small.
To reveal this gap in the uncertain case by an analytical approach is left
for future research.
\end{exmp}

\section{Conclusion}\label{sec,conclusion}
In this paper, we have studied the stabilization problem of uncertain systems
via data-rate constrained channels.
We have derived a necessary condition and a sufficient condition for stability
and have proposed a nonuniform quantizer which may reduce the required
data rate compared with the uniform one.
In particular, for scalar plants, the conditions are necessary
and sufficient, and the proposed quantizer minimizes the required data rate.

For future work, it is of interest to generalize the classes of plants and
controllers so that they include non-ARX forms and general causal controllers.
Furthermore, since the derived conditions contain some conservativeness for
multi-dimensional plants, we would like to find tighter bounds.


\small
\begin{thebibliography}{10}
\bibitem{Barmish1994}
B.~R. Barmish, \emph{{New Tools for Robustness of Linear Systems}}.\hskip 1em
  plus 0.5em minus 0.4em\relax Macmillan, 1994.

\bibitem{Bhattacharyya1995}
S.~P. Bhattacharyya, H.~Chapellat, and L.~H. Keel, \emph{{Robust Control: The
  Parametric Approach}}.\hskip 1em plus 0.5em minus 0.4em\relax Prentice-Hall,
  1995.

\bibitem{Cover2006}
T.~M. Cover and J.~A. Thomas, \emph{{Elements of Information Theory}},
  2nd~ed.\hskip 1em plus 0.5em minus 0.4em\relax Wiley, 2006.

\bibitem{Elia2001}
N.~Elia and S.~K. Mitter, ``{Stabilization of linear systems with limited
  information},'' \emph{\IEEEJAC}, vol.~46, no.~9, pp. 1384--1400, 2001.

\bibitem{Fu2005}
M.~Fu and L.~Xie, ``{The sector bound approach to quantized feedback
  control},'' \emph{\IEEEJAC}, vol.~50, no.~11, pp. 1698--1711, 2005.

\bibitem{Fu2010}
M.~Fu and L.~Xie, ``{Quantized feedback control for linear uncertain
  systems},'' \emph{\IJRN}, vol.~20, no.~8, pp. 843--857, 2010.

\bibitem{Hayakawa2009}
T.~Hayakawa, H.~Ishii, and K.~Tsumura, ``{Adaptive quantized control for linear
  uncertain discrete-time systems},'' \emph{Automatica}, vol.~45, no.~3, pp.
  692--700, 2009.

\bibitem{Kharitonov1979}
V.~L. Kharitonov, ``{Asymptotic stability of an equilibrium position of a
  family of linear differential equations},'' \emph{Differential Equations},
  vol.~14, pp. 1483--1485, 1979.

\bibitem{Li2004}
K.~Li and J.~Baillieul, ``{Robust quantization for digital finite communication
  bandwidth (DFCB) control},'' \emph{\IEEEJAC}, vol.~49, no.~9, pp. 1573--1584,
   2004.

\bibitem{Liberzon2005}
D.~Liberzon and J.~P. Hespanha, ``{Stabilization of nonlinear systems with
  limited information feedback},'' \emph{\IEEEJAC}, vol.~50, no.~6, pp.
  910--915, 2005.

\bibitem{Luenberger1997}
D.~G. Luenberger, \emph{{Optimization by Vector Space Methods}}.\hskip 1em plus
  0.5em minus 0.4em\relax Wiley, 1997.

\bibitem{Martins2006}
N.~C. Martins, M.~A. Dahleh, and N.~Elia, ``{Feedback stabilization of
  uncertain systems in the presence of a direct link},'' \emph{\IEEEJAC},
  vol.~51, no.~3, pp. 438--447, 2006.

\bibitem{Minero2009}
P.~Minero, M.~Franceschetti, S.~Dey, and G.~N. Nair, ``{Data rate theorem for
  stabilization over time-varying feedback channels},'' \emph{\IEEEJAC},
  vol.~54, no.~2, pp. 243--255, 2009.

\bibitem{Moore1966}
R.~E. Moore, \emph{{Interval Analysis}}.\hskip 1em plus 0.5em minus 0.4em\relax
  Prentice-Hall, 1966.

\bibitem{Nair2000}
G.~N. Nair and R.~J. Evans, ``{Stabilization with data-rate-limited feedback:
  Tightest attainable bounds},'' \emph{\SysCL}, vol.~41, no.~1, pp. 49--56,
  2000.

\bibitem{Nair2004}
G.~N. Nair and R.~J. Evans, ``{Stabilizability of stochastic linear systems
  with finite feedback data rates},'' \emph{\SIAMCO}, vol.~43, no.~2, pp.
  413--436, 2004.

\bibitem{Nair2007}
G.~N. Nair, F.~Fagnani, S.~Zampieri, and R.~J. Evans, ``{Feedback control under
  data rate constraints: An overview},'' \emph{\IEEEJPROC}, vol.~95, no.~1, pp.
  108--137, 2007.

\bibitem{Nesic2009}
D.~Ne\v{s}i\'c and D.~Liberzon, ``{A unified framework for design and analysis
  of networked and quantized control systems},'' \emph{\IEEEJAC}, vol.~54,
  no.~4, pp. 732--747, 2009.

\bibitem{Okano2012b}
K.~Okano and H.~Ishii, ``{Data rate limitations for stabilization of uncertain
  systems},'' in \emph{Proceedings of the 51st IEEE Conference on Decision and
  Control}, 2012, pp. 3286--3291.

\bibitem{Okano2012}
K.~Okano and H.~Ishii, ``{Data rate limitations for stabilization of uncertain
  systems over lossy channels},'' in \emph{Proc.\ of the 2012 American
  Control Conference}, 2012, pp. 1260--1265.

\bibitem{Phat2004}
V.~N. Phat, J.~Jiang, A.~V. Savkin, and I.~R. Petersen, ``{Robust stabilization
  of linear uncertain discrete-time systems via a limited capacity
  communication channel},'' \emph{\SysCL}, vol.~53, no.~5, pp. 347--360,
  2004.

\bibitem{Rohn1989}
J.~Rohn, ``{Systems of linear interval equations},'' \emph{\LAA}, vol. 126, pp.
  39--78, 1989.

\bibitem{Shingin2012}
H.~Shingin and Y.~Ohta, ``{Disturbance rejection with information constraints:
  Performance limitations of a scalar system for bounded and Gaussian
  disturbances},'' \emph{Automatica}, vol.~48, no.~6, pp. 1111--1116,
  2012.

\bibitem{Tatikonda2004a}
S.~Tatikonda and S.~Mitter, ``{Control over noisy channels},'' \emph{\IEEEJAC},
  vol.~49, no.~7, pp. 1196--1201, 2004.

\bibitem{Tatikonda2004}
S.~Tatikonda and S.~Mitter, ``{Control under communication constraints},''
  \emph{\IEEEJAC}, vol.~49, no.~7, pp. 1056--1068, 2004.

\bibitem{Tatikonda2004b}
S.~Tatikonda, A.~Sahai, and S.~Mitter, ``{Stochastic linear control over a
  communication channel},'' \emph{\IEEEJAC}, vol.~49, no.~9, pp. 1549--1561,
  2004.

\bibitem{Tsumura2009a}
K.~Tsumura, ``{Optimal quantization of signals for system identification},''
  \emph{\IEEEJAC}, vol.~54, no.~12, pp. 2909--2915, 2009.

\bibitem{Tsumura2009}
K.~Tsumura, H.~Ishii, and H.~Hoshina, ``{Tradeoffs between quantization and
  packet loss in networked control of linear systems},'' \emph{Automatica},
  vol.~45, no.~12, pp. 2963--2970, 2009.

\bibitem{Wong1999}
W.~S. Wong and R.~W. Brockett, ``{Systems with finite communication bandwidth
  constraints II: Stabilization with limited information feedback},''
  \emph{\IEEEJAC}, vol.~44, no.~5, pp. 1049--1053, 1999.

\bibitem{You2010}
K.~You and L.~Xie, ``{Minimum data rate for mean square stabilization of
  discrete LTI systems over lossy channels},'' \emph{\IEEEJAC}, vol.~55,
  no.~10, pp. 2373--2378, 2010.

\end{thebibliography}
\normalsize
\end{document}